\documentclass[12pt]{iopart}

\usepackage{iopams}
\usepackage{graphicx}
\usepackage{dcolumn}
\usepackage{bm}
\usepackage{epsfig}
\usepackage{psfrag}

\begin{document}

\title{Discrete Fracture Model with Anisotropic Load Sharing}
\author{Ra\'ul Cruz Hidalgo} 
\address{AMADE, Departament de F\'{\i}sica,  Departament de Enginyeria Mec\`anica
i de la Construcci\'o Industrial, Universitat de Girona 
Av. Montilivi s/n,  17071-Girona, Spain \\
}

\author{Stefano Zapperi}
\address{CNR-INFM, SMC, Department of Physics,
Sapienza --- Universit\`a di Roma, Piazzale Aldo Moro 2, 00185 Roma, Italy}
\author{Hans J. Herrmann}
 \address{Computational Physics IfB, HIF, E12, ETH, 8093 Z\"urich, Switzerland}

\date{\today}

\begin{abstract}

A two-dimensional fracture model where the interaction among elements
is modeled by an anisotropic stress-transfer function is presented.
The influence of anisotropy on the macroscopic properties of the samples 
is clarified, by interpolating between several limiting cases of load sharing.
Furthermore, the critical stress and the distribution of 
failure avalanches are obtained numerically for different values
of the anisotropy parameter $\alpha$ and as a function of the 
interaction exponent $\gamma$. From numerical results, one can certainly 
conclude that the anisotropy does not change the crossover point 
$\gamma_c=2$ in 2D. Hence, in the limit of infinite system size, the crossover 
value $\gamma_c=2$  between local and global load sharing is the 
same as the one obtained  in the isotropic case. In the case of finite 
systems, however, for $\gamma\le2$,  the global load sharing behavior 
is approached very slowly. 

\end{abstract}
\maketitle

\section{Introduction}
\label{intro}

For many years, the scientific community has shown great interest in the 
fracture 
of composites materials under imposed external stresses \cite{h90,cha,alava}.  
By now several aspects of this process are well understood but a definite and 
complete physical description has not been made yet. 
Furthermore, the huge technological impact of composites materials has led 
to a continuous  
development of new models and theoretical approaches \cite{h90,cha,alava}.
In fracture mechanics, the average mechanical properties of the specimen 
are commonly considered to be the input data for material modeling.  
\cite{h90,cha,alava}. Nevertheless, heterogeneous materials, such as 
 fiber-reinforced 
composites, present widely distributed local mechanical properties. Thus, 
analytical approaches are very limited and numerical simulations have become 
an indispensable tool in this field. On the other hand, the latest 
developments in statistical mechanics have led  to a deeper understanding
 of breakdown phenomena in heterogeneous systems \cite{h90,cha,alava}.  

Fiber reinforced composite materials exhibit a large variability of ultimate 
macrosopic properties. Heterogeneity and anisotropy, in the 
micro- meso- and macro-structure of the  composite, result in a complex  
scenario of damage mechanisms. Basically, the damage mechanisms include 
fiber breakage, matrix cracking and yielding, fiber-matrix  debonding and 
delamination \cite{bax95,bey97b,ibn97a,pho00}. 
In this framework, the simulation of the composite behavior  may be 
achieved by the statistical modeling of the micro-structure and the 
development of the relation between 
micro-structure and macro-behavior \cite{bax95,bey97b,ibn97a,pho00}. 

Until now the modeling of the fracture of laminar composites 
has been based on finite element  calculations $FEM$ and some  
micro-mechanical models \cite{Tsai,anand,drapier}. This modeling, 
has clarified that under  pure shear loading the overall response 
of the sample is controlled mainly by the resin response. 
In fact, the scientific community recognizes 
that $FEM$ techniques provide an excellent tool for 
predicting composite performance in controlled loading conditions. 
However, the continuous nature  of $FEM$ models usually makes them 
unable to  describe the local damage evolution; which is the 
primary micromechanical process. Therefore, a local approach is mandatory 
for fully understanding this complicated process, from physics viewpoint.

During the last two decades  Monte Carlo simulations have been used to 
numerically study 
stress redistribution in $2D$ and $3D$ for different fiber arrangements 
\cite{bax95,bey97b,ibn97a,pho00}. As a results, several aspects of composite 
fracture, when the external load is parallel to the 
direction of the fibers, have been clarified. Nevertheless, the fracture 
process and damage evolution  of anisotropic systems,  such as laminar 
composite materials subjected to  shear external stress, are far from being 
well understood.

A very useful approach to the fracture problem are the 
well-known Fiber Bundle Models (FBM), introduced long 
time ago  by Daniels \cite{daniels45} and Coleman \cite{col57} and 
subjects of intense research during the last years
\cite{zape97,new95,kun00,raul_yamir,raul_ferenc,prad1,prad2,sor92,curtin98,hansen94,hemmer92,harlow85,klo97}.  
FBMs are constructed so that a set of fibers is arranged in parallel, 
with each one having
a statistically distributed strength. 
The specimen is loaded parallel to the  fiber direction and the fibers 
break if the 
load acting on them exceeds their threshold value. Once the fibers begin 
to fail, 
several load transfer rules can be chosen. 
The complex evolution of internal damage and its associated stress 
redistribution are the most important factors to take into account 
in the accurate prediction of material strengths. 
The simplest case is to assume global load sharing $GLS$, which means that 
after each fiber
failure, its load is equally redistributed among all the intact fibers
remaining in the set. Otherwise, in local load sharing $LLS$ the overload 
is only transferred to the nearest neighbors. This case represents short 
range interactions among the fibers. However, in actual heterogeneous 
materials 
stress redistribution should fall somewhere between $LLS$ and $GLS$. 

In this paper, a generalized discrete model, where the 
interaction among  elements is described by an anisotropic 
stress-transfer function, is introduced. By varying the anisotropy strength 
and the effective range of  interaction we interpolate between several 
limiting cases of load sharing. The work is organized as 
follows. In section \ref{themodel} the model and the way in which simulations 
are carried out 
are explained in detail.  Numerical results are presented and 
discussed  in section \ref{section3}. The conclusions are given in 
the final section.

\section{The Model}
\label{themodel}
The fracture of fiber-reinforced composites is characterized by a highly
localized concentration of stresses at initial cracks. Anisotropic 
laminar reinforcement prevents the nucleation of small cracks and 
the propagation of damage. In this way, the final collapse of small 
cracks in a critical cluster is avoided, retarding sample failure.  

In materials science, the Weibull distribution \cite{cha} 
has proved to be a good empirical statistical distribution for representing sample 
strengths,
$P(\sigma)=1-e^{-(\frac{\sigma}{\sigma_{o}})^{\rho}}$. $\rho$ is 
the so-called Weibull index, which controls the degree
of threshold disorder in the system (the bigger the Weibull index, the
narrower the range of threshold values), and $\sigma_{o}$ is a
reference load which acts as unity. 
On the other hand, in continuous homogeneous materials, the load profiles 
around a local damage area can be well fitted by a power law, 
\begin{equation}
\Delta \sigma \sim r_{ij}^{-\gamma},
\label{eq4}
\end{equation}
where $\Delta \sigma$ is the stress increase on a material element at a
distance $r$ from the crack tip. The above general relation covers 
the cases of global and local load sharing,  $\gamma \rightarrow 0$, and 
$\gamma \rightarrow \infty$, respectively. The transition from these 
limiting cases has  been successfully described in isotropic 
systems \cite{raul_yamir}. In global load sharing approach, the strength of the 
sample can be computed analytically as 
$\sigma_{GLS}=\frac{\sigma_c}{\sigma_o}=(\rho e)^{-1/\rho}$  
for a Weibull distribution and $\sigma_{GLS}=\frac{\sigma_c}{\sigma_o}=\frac{1}{4}$ 
for a uniform distribution $P(\sigma)=\sigma/\sigma_o$, of breaking thresholds.  

Starting from these results \cite{raul_yamir} a discrete 
model with anisotropic load sharing is introduced. 
In the model, it is assumed a two-dimensional 
square-lattice of $N$ elements, 
each one having a strength taken from a given cumulative distribution 
$P(\sigma)$ and identified 
by an integer $i$, $1\leq i\leq N$.  Thus, to each element $i$, a random threshold 
value $\sigma_{i_{th}}$ is assigned.  

The system is driven by quasi-statically increasing the load on each 
element. The element $i$ breaks when its stress $\sigma_i$ is equal to its 
threshold value $\sigma_{i_{th}}$. Hence, the minimum value of  
$\sigma_k-\sigma_{k_{th}}$ in the set $I$ of all unbroken elements, 
\begin{equation}
\delta \sigma_i^{min} = \min\limits_{ {\displaystyle{k\in I} }} [\sigma_k-\sigma_{k_{th}}]
\end{equation}
defines the load increment $\delta \sigma_i^{min}$. 
The quasi-statically load increasing is then performed 
by adding the amount of load $\delta \sigma_i^{min}$ to all the intact elements in the system. 
Following this approach,  all intact elements have a nonzero probability
of being affected by the ongoing failure event, and the
additional load received by an intact element $i$ depends on its
$\Delta x_{ij}$ and $\Delta y_{ij}$ from the element $j$ which has just been broken. 
Furthermore, an anisotropic interaction 
is assumed between elements such that the load received by an
element $j$, due to the failure of $i$, follows the relation: 
\begin{equation}
F(\Delta x_{ij},\Delta y_{ij},\gamma,\alpha) = Z_i \left( \alpha{\Delta x_{ij}}^{2}  +  (1-\alpha){\Delta y_{ij}}^{2}\right)^{-\gamma}, 
\label{eq5}
\end{equation}
where $\Delta x_{ij}$ and $\Delta y_{ij}$ are their relative distances 
on the $x$-axis and $y$-axis, respectively. $\gamma$ and $\alpha$ are adjustable parameters and 
the value $Z_i$ is always given by the normalization condition, 
\begin{equation}
Z_i = 1/\sum_{j\in I} \left( \alpha{\Delta x_{ij}}^{2}  +  (1-\alpha){\Delta y_{ij}}^{2}\right)^{-\gamma}, 
\label{eqz}
\end{equation}
The sum runs over the set $I$ of all unbroken elements. We assume periodic boundary conditions, 
which means the largest value of $\Delta x_{ij}$ and $\Delta y_{ij}$ is  $\frac{L-1}{2}$, where $L$ is 
the linear size of the system. We note here that the assumption of periodic boundary conditions 
is for simplicity. In principle, an Ewald summation procedure would be more accurate.  In Eq.(\ref{eq5})
 the extreme cases $\gamma\rightarrow 0$  and $\gamma\rightarrow\infty$ 
also correspond to global load sharing and local load 
sharing, respectively. Strictly speaking, for all $\alpha$, the range of 
interaction covers the whole lattice. When changing the anisotropy 
factor $\alpha$, one moves  from a completely  anisotropic case (either $\alpha=0$ or $\alpha=1$) 
to the isotropic load redistribution ($\alpha=0.5$). 

Following this approach, a failing element transfers its load to the surviving 
elements of the set. This may provoke secondary fractures in the system 
which in turn induce tertiary
ruptures and so on until the system fails or reaches an equilibrium state
where the load on the intact elements is lower than their individual
strength. In this latter case, the external force is
increased again and the process is repeated until the 
material macroscopically fails. The size of an avalanche $\Delta$ is defined as
the number of broken elements between two
successive external drivings. Hence, during an avalanche 
of failure events, an intact element $i$ receives
 the excess load from failing elements
$j$ at each time step. Consequently, its load increases by an amount,
\begin{equation}
\sigma_i(t+\tau)=\sum\limits_{j\in
  B(\tau)}\sigma_{j}(t+\tau-1)F(r_{ij},\gamma,\alpha), 
\label{eq6}
\end{equation}
where the sum runs over the set $B(\tau)$ of elements that have failed
at time step $\tau$. Thus,
$\sigma_i(t_0+T)=\sum\limits_{\tau=1}^{T}\sigma_i(t_0+\tau)$ is the
total load element $i$ receives during an avalanche initiated at $t_0$
and finished at $t_0+T$. In this way, when an avalanche ends, the
external load is increased again and another avalanche is
initiated. The process is repeated until no intact elements remain in
the system and the ultimate strength of the material $\sigma_c$, is
defined as the maximum load the system can support before its complete
breakdown.

\section{Simulation Results}
\label{section3}

\begin{figure}[b]
\begin{center}
\includegraphics[angle=-90,width=7.7cm]{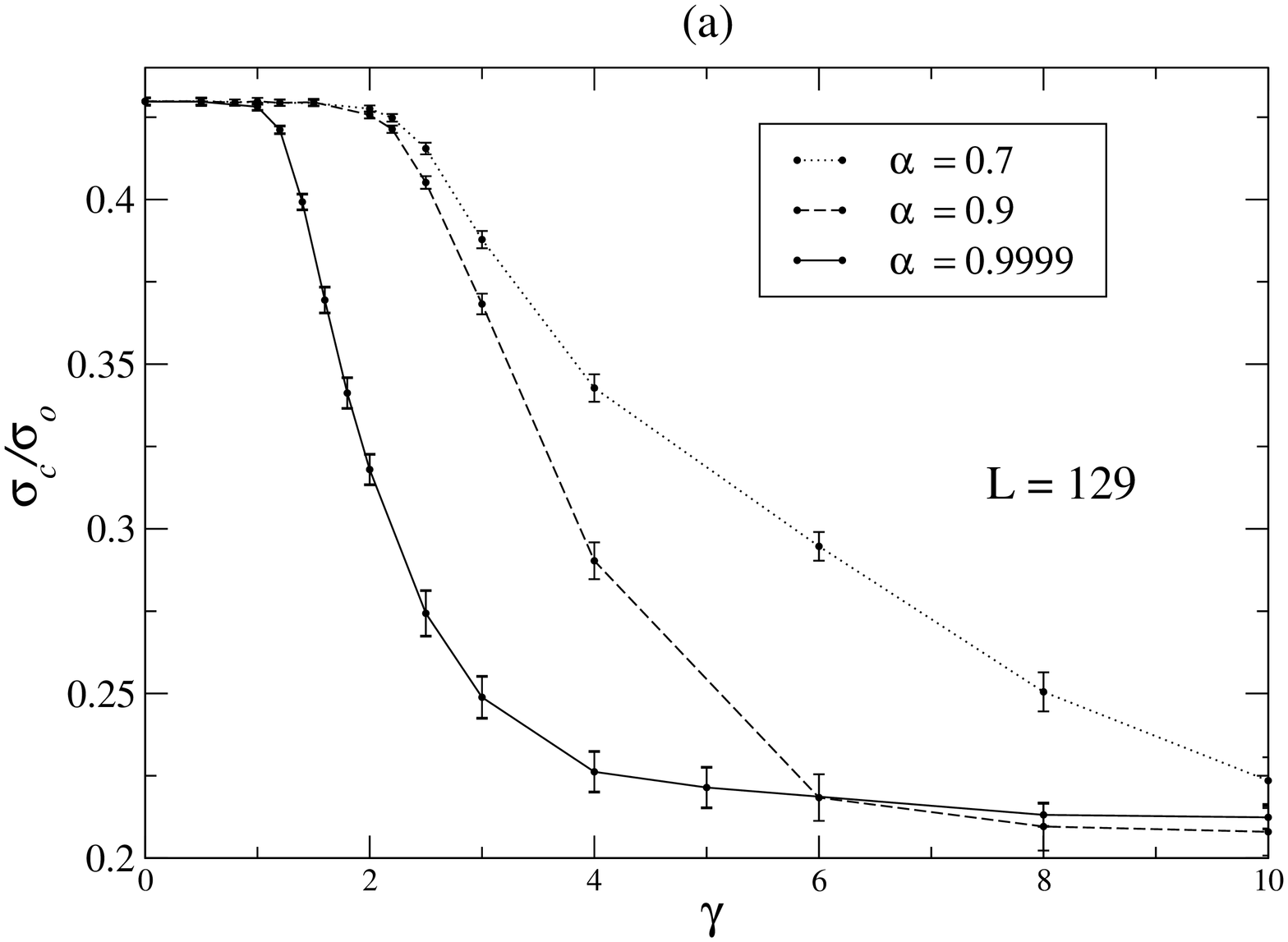}
\includegraphics[angle=-90,width=7.7cm]{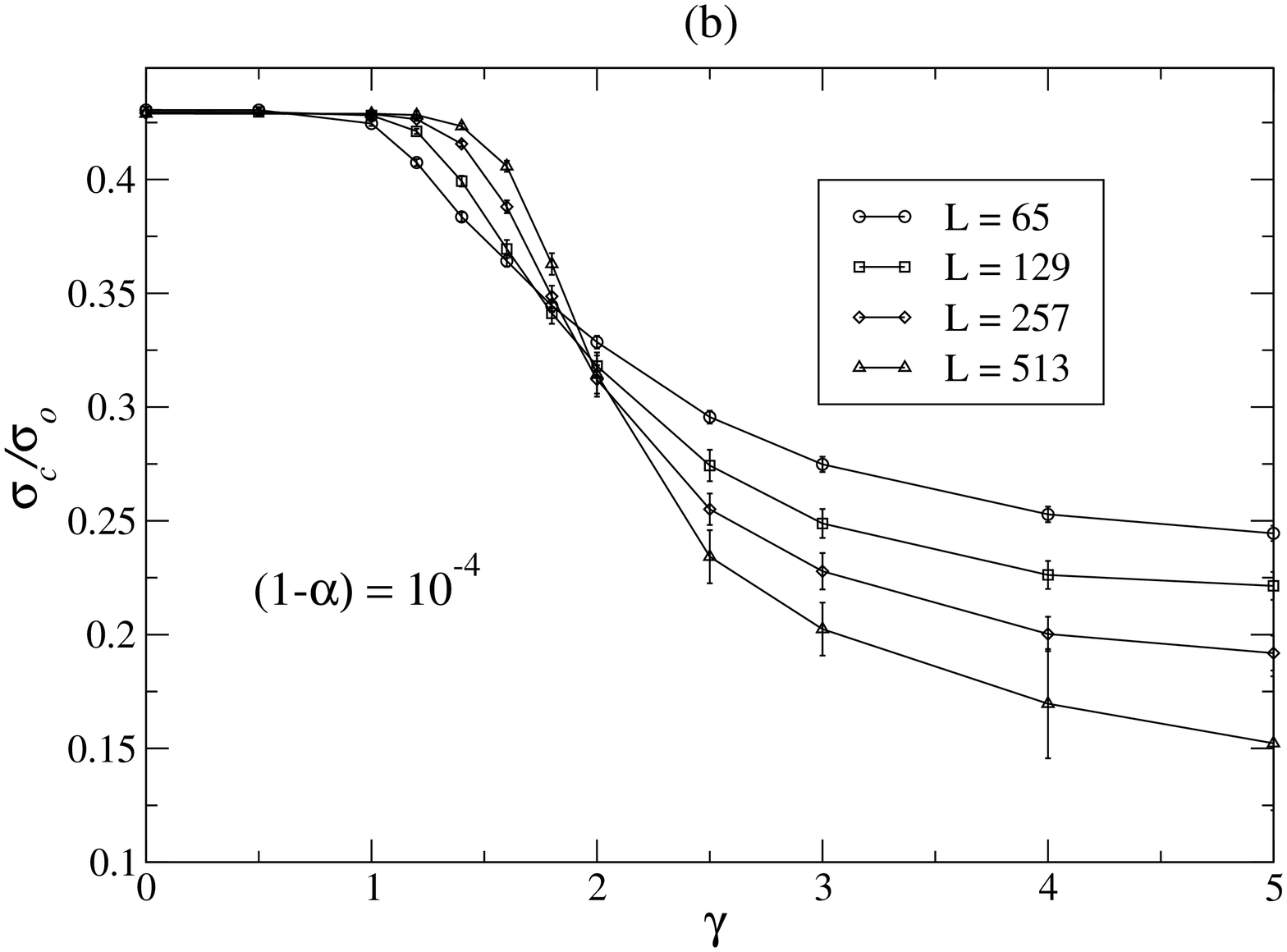}
\caption{
The ultimate strength $\frac{\sigma_c}{\sigma_o}$ 
of the system is studied, for a Weibull distribution  of thresholds with  $\rho=2$ and $\sigma_{0}=1$. 
a) We explore different ranges of  interaction $\gamma$ and anisotropy $\alpha$.  
b) Results for  different system sizes $L$ and ranges of interaction $\gamma$ are presented}
\label{figure1}
\end{center}
\end{figure}

The mechanical properties of the bundle and the statistics of internal damage events  
were studied  numerically by varying the anisotropy strength ($\alpha$) and the range 
of interaction ($\gamma$). Large scale numerical simulations in $2D$ were executed.  Several system 
sizes $(L=65,129,257,513,1025)$ were considered and simulations  were performed  over at least five 
different realizations for the biggest system $L=1025$ and five thousand for the smallest one $L=65$. We recorded the avalanche size distribution $D(\Delta)$, and the ultimate strength of the samples $\frac{\sigma_c}{\sigma_o}$, which is always normalized by 
the characteristic value $\sigma_o$ of the cumulative threshold distribution 
$P(\sigma)=P(\frac{\sigma}{\sigma_o})$ . 

It is important to remark, that by using an isotropic power law stress 
redistribution $\Delta \sigma_{add} \sim r^{-\gamma}$, a crossover point was 
observed \cite{raul_yamir}.  Hence, two distinct regions were distinguished, 
over the domain of $\gamma$. For small $\gamma$, $\frac{\sigma_c}{\sigma_o}$ 
is independent on $L$, which corresponds to the GLS behavior. However, when the effective 
range of interaction is decreased $\gamma > \gamma_{c}$,  the limiting case of local 
load sharing is approached and the  strength of the system should vanish in the 
$L\to\infty$ limit \cite{harlow85,klo97}.  In the isotropic case, 
$\gamma_c$  falls in the vicinity of $\gamma_c=2$  \cite{raul_yamir}.

Figures \ref{figure1}a. and \ref{figure1}b show the critical stress values $\frac{\sigma_c}{\sigma_o}$ 
obtained for several ranges of interactions  $\gamma$  and anisotropy strengths 
$\alpha$.  The data corresponds to systems with $129\times129$ elements and 
a  Weibull  distribution of breaking thresholds with $\rho=2$ and $\sigma_o=1$. 
Two distinct regions can be easily identified in the graph \ref{figure1}a.  
For small $\gamma$, the critical strength $\frac{\sigma_c}{\sigma_o}$ is independent, 
within  statistical errors, of both the effective range of interaction $\gamma$ and 
the anisotropy  strength $\alpha$. As we have already pointed 
out, this behavior is expected  for the standard $GLS$ scheme. However, when the 
effective range of interaction decreases or the anisotropy  strength increases, 
we found non-trivial dependencies. In Figure\ \ref{figure1}b, the critical 
stress of a system with strong anisotropy 
($\alpha \approx 0.9999$) is detailed. 
We illustrate the model's behavior for several system sizes from $L=65$ to $L=513$. 
For small $\gamma$,  $\frac{\sigma_c}{\sigma_o}$ is independent of both 
effective range of interaction and system size. However, as soon as the localized 
nature of  the  interaction becomes dominant, {\em i.e.} $\gamma>\gamma_c$,
$\frac{\sigma_c}{\sigma_o}$  vanishes logarithmically with increasing system size.  
This qualifies for a genuine short-range behavior 
as found in local load sharing models, where  the strength of the sample must vanish 
in the thermodynamic limit \cite{harlow85,klo97}. 
In summary, in the regime 
$\gamma \leq 1$, all the numerical findings are in 
excellent agreement with the mean field analytical prediction. 
Besides, figure \ref{figure1}a suggests that the crossover point $\gamma_{c}$  would differ 
from what was found using the isotropic stress transfer function  \cite{raul_yamir} only for a 
very high anisotropy strength, {\em i.e.} $\alpha > 0.9999$.

Nevertheless, a priori one would expect the system size dependence to be more  pronounced 
onces the anisotropy in introduced. Figure \ref{figure1}b shows that in the transition  
region the system size dependence is more appreciable than for the isotropic 
case \cite{raul_yamir}.  On the left side of the transition region the critical  
stress  $\frac{\sigma_c}{\sigma_o}$  slowly increases with increasing system size. 
Note that for the isotropic case  the convergence to the thermodynamic limit is faster; 
consequently, a more accurate estimation of the critical point $\gamma_c=2$ could be done 
\cite{raul_yamir}. In the present case, in order to find a reasonable estimation of $\gamma_c$, 
we used the fact that on the short range interaction region ($LLS$) the convergence to the 
thermodynamic limit is qualitatively different than on the long range interaction region ($GLS$). 
On the $GLS$ side of the transition region the critical  stress  $\frac{\sigma_c}{\sigma_o}$  
increases with increasing system size, towards the $GLS$ exact solution. Contrary, on the $LLS$ 
side of the transition region $\frac{\sigma_c}{\sigma_o}$ vanishes logarithmically with 
increasing system size. Despite the fact that figure \ref{figure1}a seems to indicate 
$\gamma_{c} \neq 2$ for $\alpha = 0.9999$, figure \ref{figure1}b also suggests that for any 
value $\alpha<1$, the critical value $\gamma_{c}$ shifts towards $\gamma_{c}=2$ as the 
thermodynamic limit is reached. For elucidating if the value of $\gamma_{c}$ changes onces 
the anisotropy is introduced we then focused on $\gamma=1$ and $\gamma=2$. Hence, 
changing the anisotropy strength $\alpha$ and studying the convergence of 
$\frac{\sigma_c}{\sigma_o}$ to the thermodynamic limit, a better estimation of $\gamma_{c}$ is done. 

\begin{figure}[b]
\begin{center} 
\includegraphics[angle=-90,width=7.7cm]{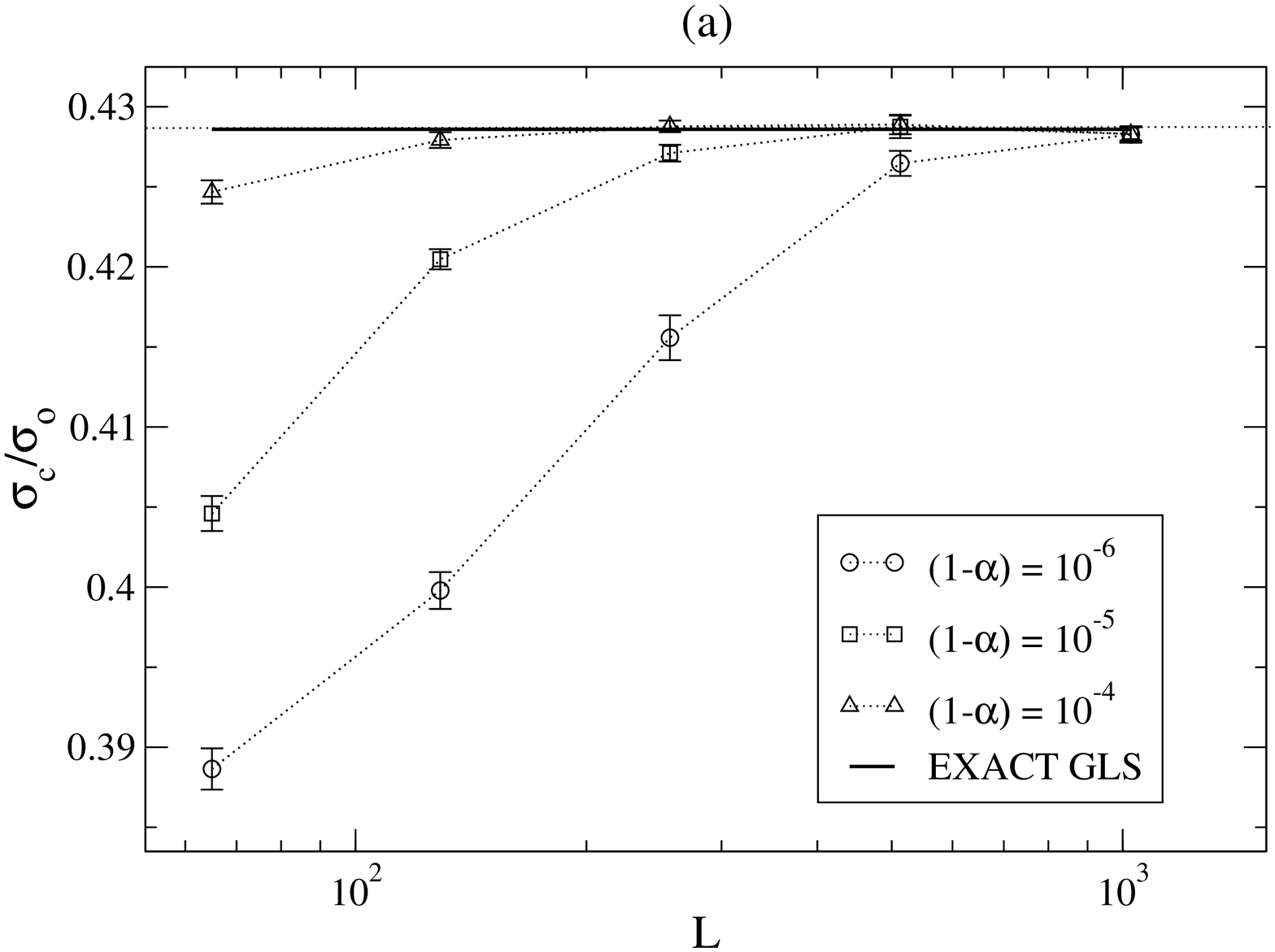}
\includegraphics[angle=-90,width=7.7cm]{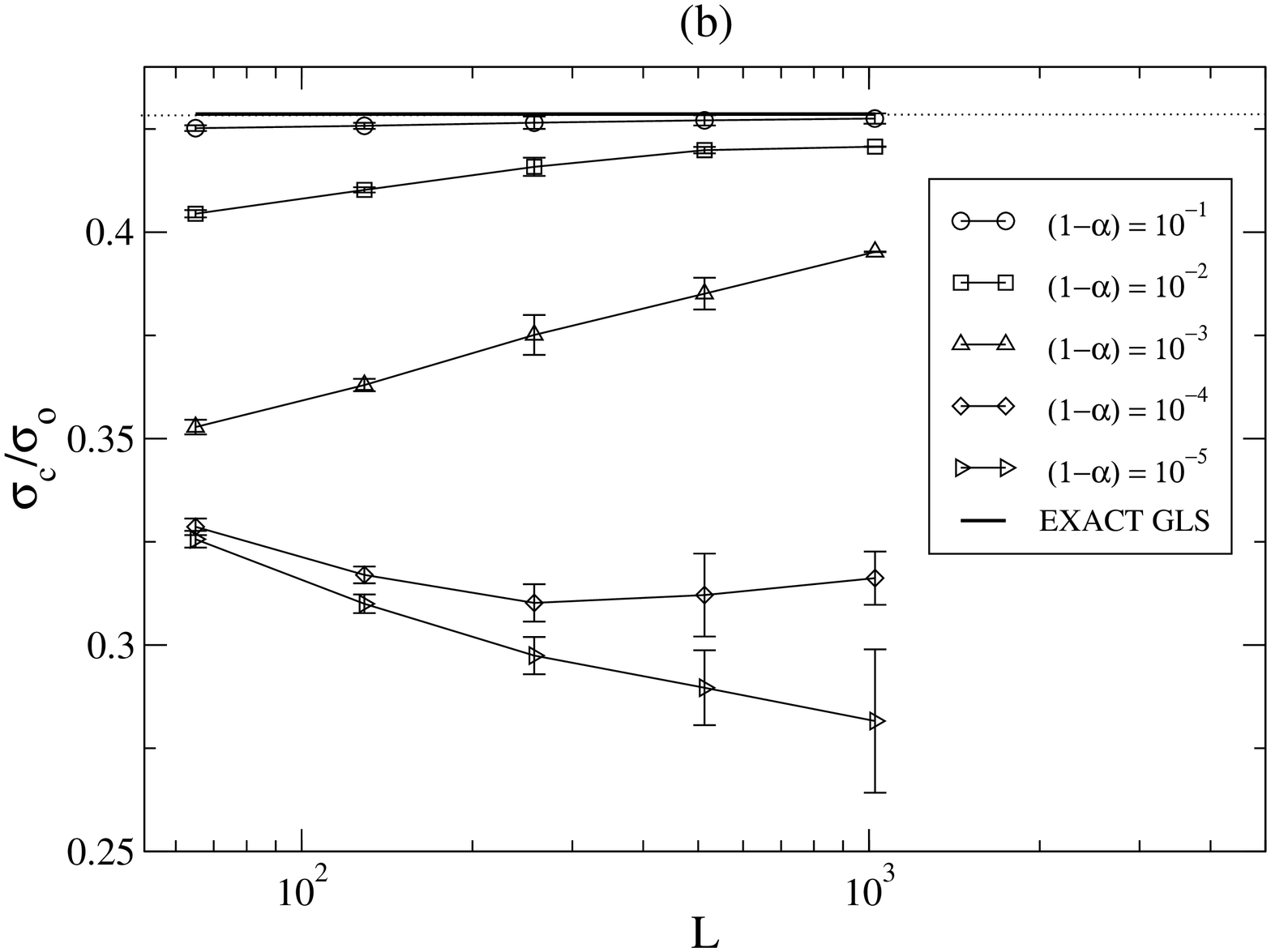}
\caption{We show results of ultimate strength as a function of the system size, 
for several anisotropy  strengths. A Weibull distribution  of thresholds with  $\rho=2$ and 
$\sigma_{0}=1$ is used.  a) results obtained at $\gamma=1$, b) results obtained at $\gamma=2$}
\label{figure2}
\end{center}
\end{figure}

In Figure \ref{figure2}, we describe the size dependency of the ultimate strength 
$\frac{\sigma_c}{\sigma_o}$ for a  Weibull distribution of thresholds with $\rho=2$ 
and $\sigma_o=1$. Figure \ref{figure2}a and  figure \ref{figure2}b illustrate the outcomes at 
$\gamma=1$ and $\gamma=2$, respectively.  Several anisotropy  strengths were explored and it 
was found that when the system size is increased,  the anisotropy plays a weaker role.  
Moreover, as expected  the system size dependence is more pronounced for $\gamma=2$ than for $\gamma=1$.  
We note in Fig.\ref{figure2}b that numerical uncertainties surface when one gets numerically very 
close to $\alpha=1$, for small system sizes. That is due to the fact that in a $2D$ square lattice 
topology as $\alpha \rightarrow 1$, the load sharing from one row to the next might become so small 
that the rows would  become effectively decoupled. That is certainly an undesirable topology effect 
which might also magnify the system size effects. However, our results indicate that even in the 
presence of high anisotropy, at $\gamma_c=2$, the system shows a tendency to behave as $GLS$ when 
approaching to the thermodynamic limit, within our numerical uncertainties. 
Consequently, the crossover point for the anisotropic variable range of interaction given  
by Eq.(\ref{eq5}) results at $\gamma_c=2$.
  
To test the universality of this statement, we did calculations for several threshold distributions.  
A Weibull distribution with  $\rho=10$, as well as a uniform distribution were used for comparison.  
Moreover, to access accurately the crossover point, several system sizes were considered. 
In every case, we changed $\alpha$ using the values ($0.9; 0.99; 0.999$ and  $0.9999$). 
Our aim is to elucidate how far from $GLS$ behavior the system is for $\gamma=2$, 
after the anisotropy is introduced. As we pointed out before, this value $\gamma_c=2$ 
defines the crossover between $GLS$ and $LLS$ behaviors, for the isotropic case.  

\begin{figure}
\begin{center}
\includegraphics[angle=-90,width=12cm]{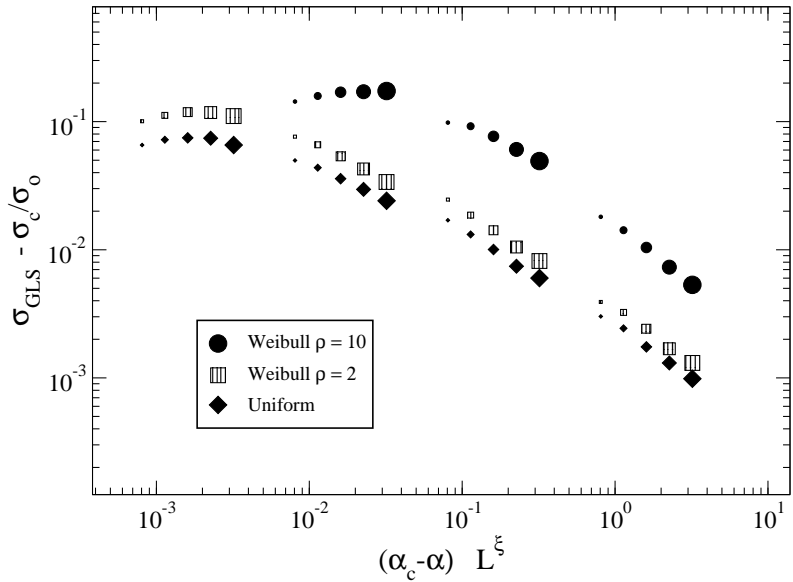}
\caption{The ultimate strength of the system $\sigma_c$ is studied at $\gamma=2$,  
through the magnitude  $\delta \sigma = \sigma_{GLS} - \frac{\sigma}{\sigma_o}$.  Several anisotropy 
strength values and $(L=65,129,257,513,1025)$ are illustrated. 
}
\label{figure3}
\end{center}
\end{figure}

In Figure\ \ref{figure3},  results of the ultimate strength of samples with 
strong anisotropy are shown in detail. The scaled magnitude 
$x = (\alpha_c-\alpha)L^\xi$  is plotted against the "distance"  
$\delta \sigma = \sigma_{GLS} - \frac{\sigma_c}{\sigma_o}$  from the 
well-defined $GLS$'s behavior. Note that each symbol ({\em i.e.} square, circle and diamond) 
is related to a different threshold distribution. In addition, the sizes of the
symbols are linked to  different system sizes (the bigger the symbol, the larger the system size). 
Different regions of the plot, illustrated with  the same symbol in five 
different sizes, correspond to a given value of $\alpha$ (from left to the right $0.9; 0.99; 0.999$ and  $0.9999$).
In the plots the data, corresponding to each threshold distribution, are aligned finding data collapse,
\begin{equation}
\delta \sigma = \left(\sigma_{GLS} - \frac{\sigma_c}{\sigma_o}\right) \sim  F( (\alpha_c-\alpha) L^\xi ) \sim F(x)
\end{equation} 
where we have introduced the scaling function $F( (\alpha_c-\alpha) L^\xi)$, 
with $\xi=0.5$ and $\alpha_c=1$.  
The results for a uniform 
distribution and  Weibull with $\rho=2$ are very similar. Furthermore, decreasing the 
disorder, {\em e.g.} Weibull  with $\rho=10$ and $\sigma_o=1$, 
only magnifies the topology and finite size effects. The scaling exponent  
$\xi \approx 0.5$ is universal, defining a new crossover exponent in $\alpha$
and $L$. From those results, one can conclude that the anisotropy 
does not change the crossover point for the range of interaction in 2D, $\gamma_c=2$. 
For finite systems, the global behavior is reached very slowly, as we get far from  
$\alpha_c=1$. We also notice that the distance to the $GLS$ behavior 
$\delta \sigma = (\sigma_{GLS} - \frac{\sigma_c}{\sigma_o})$  does vanish in the 
infinite system size limit, as a power law $\delta \sigma \sim L^{-\beta}$, 
and we estimate $\beta = 0.37\pm0.03$. We conclude, that even in presence of high anisotropy  
the behavior of the system for $\gamma\le\gamma_c=2$ shows a tendency to $GLS$ as $L\rightarrow\infty$. 
The infinite system would display a qualitatively different behavior only for $\alpha_c=1$, where 
it would become effectively a $1D$ model, with $\gamma_c=1$. 

The fracture process can also be described by the precursory activity 
before complete breakdown. The statistical properties of rupture sequences are 
characterized by the avalanche size distribution. From an experimental point of view  the 
precursory activity  
is related to the acoustic emissions generated during the fracture of
materials \cite{gar97,gar98,maes98,pet94}. The avalanche size distribution is a measure of causally 
connected broken sites. All the intact 
elements have a non-zero chance to fail independently of the (spatial) rupture history, 
and any given element could be near to its rupture point regardless of its position in the lattice.  

\begin{figure}
\psfrag{AAA}{{\LARGE $\Delta^{-\frac{5}{2}}$}}
\begin{center}
\includegraphics[angle=-90,width=12cm]{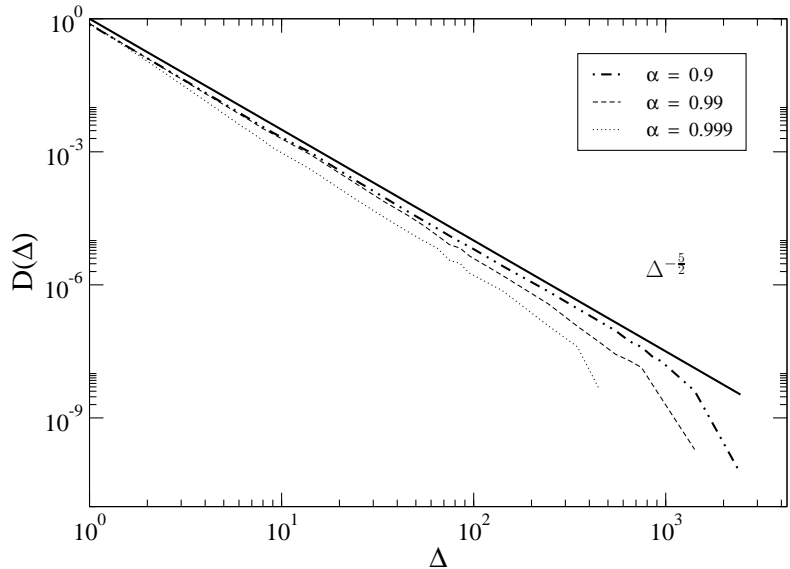} 
\caption{ Avalanche size distribution for $N=257 \times 257$ fibers. Results corresponding to several anisotropic strengths are illustrated} 
\label{figure4}
\end{center}
\end{figure}

The highly fluctuating activity is certainly related to the long range interactions where   
the avalanche size distribution can usually be well fitted by a power law $P(\Delta)\sim \Delta^{-\frac{5}{2}}$.
This actually corresponds to the mean field scenario, $GLS$ \cite{hansen94,hemmer92,harlow85,klo97}.
However, when the spatial correlations are important $LLS$, stress concentration   
takes place in the elements located at the perimeter of an already formed cluster. 
Hence, elements far away from the clusters of broken elements have significantly lower 
stresses and thus the size of the largest avalanche is reduced as well as the number of failed
elements belonging to the same avalanche, leading to lower precursory
activity. 

\begin{figure}
\begin{center}
\includegraphics[angle=-90,width=12cm]{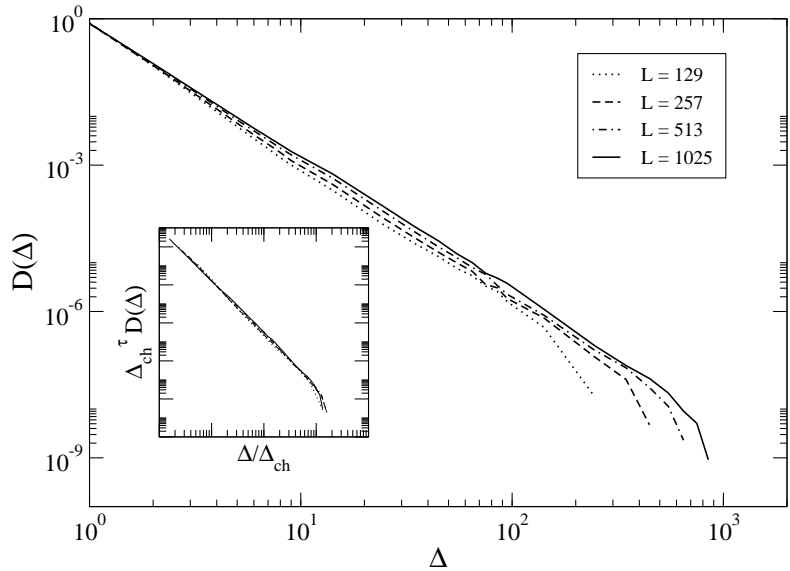}
\caption{The normalized distributions $D(\Delta)$ of avalanches for $\alpha=0.999$ are illustrated, 
several system sizes are considered. In the inset we verify the scaling ansatz given by Eq.(\ref{fun}) 
with $\Delta_{ch} \sim L^\theta$.}
\label{figure5}
\end{center}
\end{figure}

Figure\ \ref{figure4}  illustrates the avalanche statistics obtained for systems 
with $N=257 \times 257$ fibers. In every case, we have set $\gamma=2$ and used different values of 
the anisotropy strength $\alpha$.  It is noticeable that the avalanche size distributions can always 
be fitted to a power law with a non-trivial  exponent. However, as we get far from $\alpha=1$ 
the exponent tends asymptotically  to the value $\tau=\frac{5}{2}$, reflecting the tendency to 
recover the GLS's behavior. 

To characterize the system size dependence, we propose the following scaling ansatz for the avalanche 
size  distribution,
\begin{equation}
D(\Delta,L) = \Delta_{ch}^{-\tau} \,~ g(\frac{\Delta}{\Delta_{ch}})
\label{fun}
\end{equation}
where  $\Delta_{ch}$ is a characteristic avalanche $\Delta_{ch} \sim L^\theta$ 
and $g(x)$ is a scaling function that goes like $g(x)=x^{-\tau}$ for $x<1$ and decays faster than a power law for $x>1$. 
In Figure\ \ref{figure5}, the avalanche statistics resulting from 
systems with different sizes are shown.  The model is again investigated at $\gamma=2$, 
and very strong anisotropy $\alpha=0.999$.  The scaled function is presented in the inset. 
It can be seen that the data collapse yields $\theta=0.6\pm0.1$ and  
$\tau=2.50\pm0.02$ with a power law over several orders of magnitude in 
$\frac{\Delta}{\Delta_{ch}}$. This result suggests that  for $\gamma=2$, even in presence 
of very high anisotropy, the avalanches are distributed as in the case of $GLS$, namely as
$D(\Delta)\sim \Delta^{-\frac{5}{2}}$. Thus, the crossover point is still at
$\gamma_c=2$, even in the case of an anisotropic stress redistribution (given by Eq.(\ref{eq5})).  

\section{Discussion}

Long-range fiber bundle models can be considered as a first approximation to model 
the fracture behavior of a disordered elastic medium. It has been proposed in several
instances to replace the full solution of the elastic equations by 
a Green function
\cite{barthelemy02,toussaint05}. This
method has the advantage to avoid the computational cost involved in the inversion
of the elastic equations, but in principle it is only accurate for diluted damage.
A particularly simple example is provided by the random fuse model (RFM) in which a lattice
of conducting bonds with random failure thresholds is loaded by applying an external
voltage at the two ends of the lattice. When a fuse fails the current is redistributed
to the neighboring fuses by solving the Kirchhoff equations. When only a few bonds
are broken, the current is transferred according to the homogeneous lattice Green
function which  is given by $F(r)=x/r^3$ in two dimensions \cite{barthelemy02}. It is therefore a long-range
(with $\gamma=2$) and anisotropic load transfer function, but of a different character
than the one we have used here. 

\begin{figure}
\begin{center}
\includegraphics[angle=-90,width=12cm]{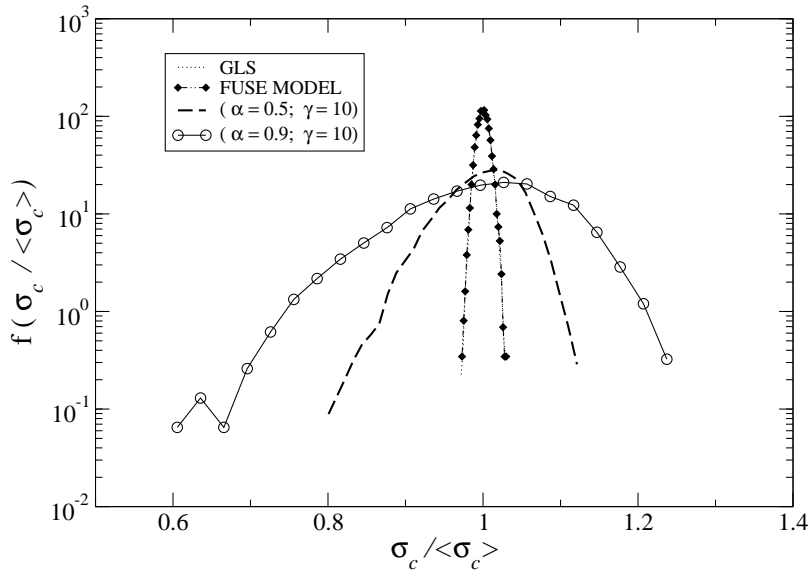}
\caption{Global strength distribution of samples with several types of load sharing ($L=129$, averaged over 5000 different configurations). $GLS$ and $RFM$ are very the similar.}
\label{figure6}
\end{center}
\end{figure}

We have simulated a fiber bundle model using a load
transfer function inspired by the RFM. The general result is that the
failure properties of our model resemble very much those of GLS
fiber bundles rather than those of the original RFM. This is particularly
apparent when looking at the strength distribution, displayed in 
Figure \ \ref{figure6}. Both the mean-field GLS approach and our result obtained with the RFM load transfer 
function obey Gaussian statistics. Notice that the original
RFM displays instead a qualitatively different log-normal strength distribution \cite{alava}.
This confirms that the Green function approach is reliable at most in
the initial stages of the damage accumulation process and it is not correct to describe
the global failure of the RFM. Figure \ref{figure6} also shows that anisotropic LLS functions result 
instead in a larger scattering of ultimate strength, and their asymmetric distributions are usually 
fitted by Weibull distribution functions.  It is noticeable from the data that introducing very 
high anisotropy strength amplifies the scattering in the global strength of the samples.  
However, the strength distribution for the anisotropic system appears to be closer to a {\it Gaussian} 
in contrast to the {\it classical} Weibull behavior, which is usually obtained for LLS approaches.

In conclusion, we have studied a discrete fracture model 
where the interaction among elements is considered to decay 
anisotropically with the 
distance from an intact element to the rupture point. 
The two classical regimes (local and global) are found as the exponent of the
stress-transfer function varies and a crossover point is again identified in the vicinity of 
$\gamma_c=2$. The strength of the material for $\gamma<\gamma_c$
does not depend on both the system size and $\gamma$ qualifying for
mean-field behavior, whereas for the short range regime, the critical
load vanishes in the thermodynamic limit.  The behavior of the model at both sides of the 
crossover point was numerically  studied by recording the avalanche and the critical 
stress for several system sizes. From our numerical results, one can certainly conclude that the 
anisotropy does not change the crossover point $\gamma_c=2$ in the 2D model, in the infinite system size limit. 
The $2D$ model would display a qualitatively different behavior only for $\alpha_c=1$, where it would become 
effectively a $1D$ model, with $\gamma_c=1$. Moreover, in finite systems for $\gamma\le2$, the global load 
sharing behavior is very slowly recovered as we get far from $\alpha_c=1$, within our numerical uncertainties.

\section{Acknowledgments}
RCH acknowledges the financial support  of the Spanish Minister of Education and Science, 
through a {\it Ramon y Cajal Program}. This work began while the authors (RCH and SZ) 
were visiting the Institute for Computational Physics (ICP) at the University of Stuttgart. 
Its financial support and hospitality are gratefully acknowledged. HJH acknowledges 
the Max Planck Prize. SZ acknowledges financial  
support from EU FP6 NEST Pathfinder programme TRIGS under contract NEST-2005-PATH-COM-043386

\end{document}